\documentclass[reprint,aps,prx,amsmath,amssymb,longbibliograaphy,doublecolumn,notitlepage,nofootinbib]{revtex4-1}
\usepackage[centering,hmargin=2.5cm,vmargin=2.5cm]{geometry}
\usepackage[utf8]{inputenc}
\usepackage{amsmath,amssymb,bm,graphicx}
\usepackage{graphics}
\usepackage{graphicx}
\usepackage[justification=centering]{caption}
\usepackage{subcaption}
\usepackage{float}
\usepackage{dblfloatfix}
 \usepackage[stable]{footmisc}
 
 \usepackage{soul}
 \usepackage{float}
 \usepackage{fixltx2e}
\usepackage{xcolor}

\usepackage[colorlinks, linkcolor= blue, citecolor = blue, urlcolor=blue]{hyperref}
\usepackage{physics}
\def\be{\begin{equation}}
\def\ee{\end{equation}}
\def \bea{\begin{eqnarray}}
\def \eea{\end{eqnarray}}

\begin{document}

\title{Study of q diagram in GX 339-4}

\author{Sanyukta Agarwal (190765)}
\email{sanyukta@iitk.ac.in}
\author{J. S. Yadav}
\email{jsyadav@iitk.ac.in}
\affiliation{Department of Physics, Indian Institute of Technology, Kanpur - 208016, India}

\begin{abstract}
Black hole X-ray binary (BHXB) GX 339-4 produces the most frequent X-ray outbursts (every 2-4 years) among known X-ray binary sources in our galaxy. Here we present the study of the evolution of GX 339-4 over the years of observations using the Hardness Intensity Diagram (HID), i.e., the q diagram. We present an analysis of two outbursts of the source observed by the Monitor of All-sky X-ray Image (MAXI) telescope aboard the International Space Station (ISS).
\end{abstract}
\maketitle

\section{Introduction}
Black Hole X-ray Binaries (BHXBs) consist of a black hole accreting material off a companion star. These are of two types, High Mass X-Ray Binaries (HMXB) and Low Mass X-Ray Binaries (LMXBs). GX339-4 is known to be a LMXB with a Blackhole mass 4-11 Solar Mass (M\textsubscript{o}) \cite{Zdziarski_2019}. GX339-4 is a transient black hole and is known to show a hysteric transition \cite{F_rst_2016}. It can be classified into four distinct states: Low-Hard State (LHS), Hard Intermediate State (HIMS), Soft Intermediate State (SIMS), and High-Soft State (HSS). LHS is primarily present in the initial and final stages of outbursts characterized with a hard spectrum and high variability and corresponds to the right branch in the HID. At high flux, the source transitions from LHS to HSS. HSS consists of a soft spectrum, low variability, and corresponds to the left part in the HID. At low flux, the sources returns to LHS. HIMS and SIMS correspond to intermediate states between these transitions, and they correspond to the middle part of the HID between HSS and LHS. HIMS appears after the initial LHS and reappears before the source returns to LHS. In the HID, it corresponds to horizontal tracks at both high and low flux. SIMS have a slightly softer energy spectrum compared to HIMS and thus are on the left of HIMS in the HID \cite{belloni}.

\begin{figure}
  \centering
  \includegraphics[height=0.6\textheight, width=.5\textwidth]{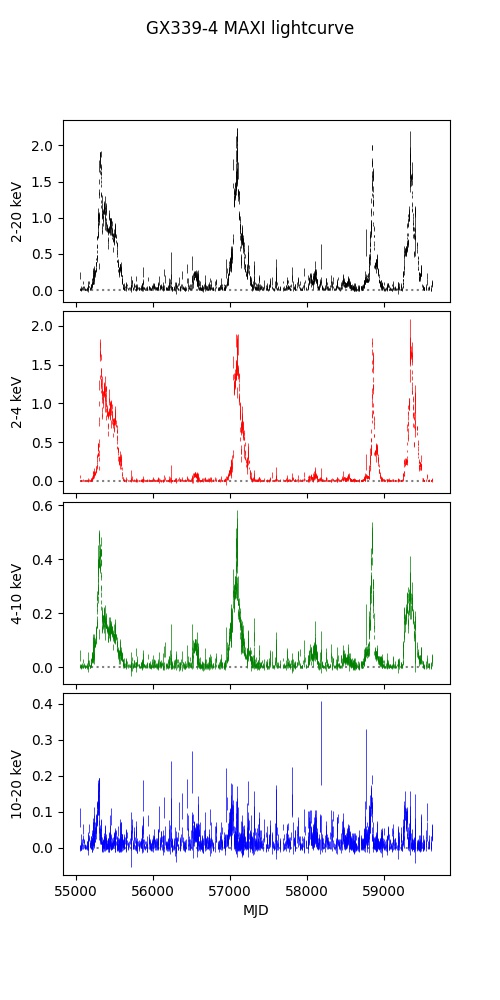}
  \caption{Background subtracted MAXI lightcurve (bin: one day) of GX339-4, modified to exclude negative flux points. MJD: Modified Julian Day on x axis and counts(flux) corresponding to 4 different energy bands on y axis.}
\label{fig:1}
\end{figure}

\section{Observation and Data Reduction}\label{sec:rules_submission}
MAXI is an X-ray telescope aboard the ISS since 2009. It performs sky surveys using a wide field view of X-ray detectors. It has an orbit of 96 minutes, thus measures the brightness of sources every 96 minutes. Developed by Japan Aerospace Exploration Agency (JAXA), it surveys the sky using the Gas Slit Camera (GSC) and Solid-state Slit Camera (SSC).\\

I used publicly available data observed mainly using GSC (operating in 2-30keV) \cite{10.1093/pasj/61.5.999} to produce the light curve. Lightcurve of bin one day (Fig. \ref{fig:1}) clearly shows four distinct outbursts in the years 2010-11, 2014-15, 2020, and 2021 respectively, from left to right. I chose to study the first-ever outburst observed by MAXI (2010-11) and compare it to the most recent outburst (2021). I had previously analyzed a few orbit data of the 2021 outburst observations by the Large Area X-ray Proportional Counter (LAXPC) instrument on board the Indian satellite AstroSat. LAXPC is a collection of three identical proportional counters (each having five layers) \cite{10.1117/12.2231857}.

\begin{figure}
    \centering
    \includegraphics[height=0.2\textheight, width=.45\textwidth]{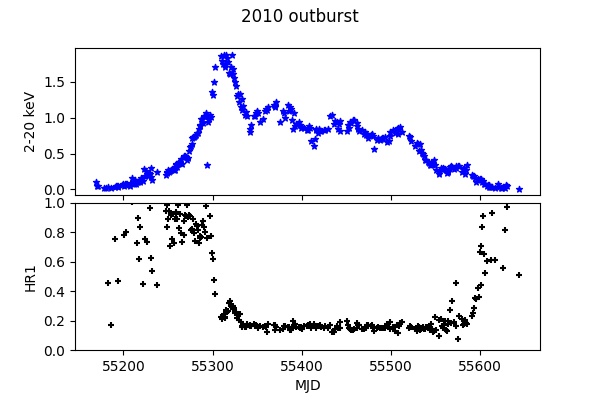}
    \includegraphics[height=0.2\textheight, width=.45\textwidth]{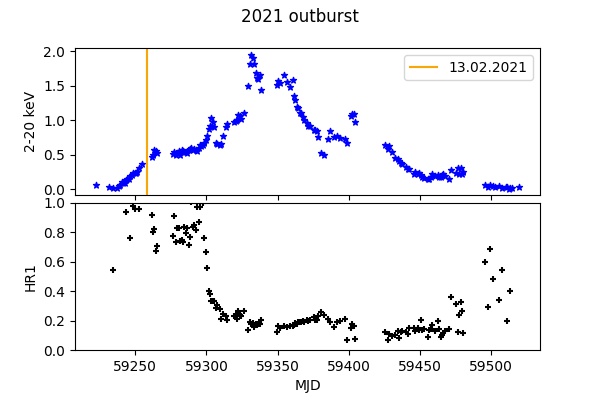}
    \caption{Background subtracted lightcurve (bin: one day) and Hardness ratio (HR1) plotted on the top and bottom panels respectively for 2010 and 2021 outburst. MJD: Modified Julian Day on x axis, counts (flux) in the 2-20 keV energy band on y axis of top panel.}
\label{fig:2}
\end{figure}    

\section{Methods and Analysis}\label{sec:types_paper}
\subsection{Lightcurve}
Publicly available background subtracted data of the source observed by MAXI contained some data points with negative flux. This is physically impossible. Thus I filtered out the negative flux data points to obtain the background subtracted lightcurve (Fig. \ref{fig:1}). The rise and gradual decrease in flux over time demarcate the start and end of an outburst. The rest of the time, when the flux count rate is nearly 0, represents the quiescent state of the transient. Comparing the four subplots in Fig. \ref{fig:1} we can infer that most of the contribution in the observation comes from light in the energy range 2-10 keV, with a lesser contribution from the energy range 10-20 keV. A significant proportion of the observation lay in the energy range 2-4 keV.\\

Across the four outbursts, the first two lasted longer than the latter two, the latter two being in close proximity to each other, whereas the first two were about four years apart.

\begin{figure}
    \centering
    \includegraphics[height=0.2\textheight, width=.45\textwidth]{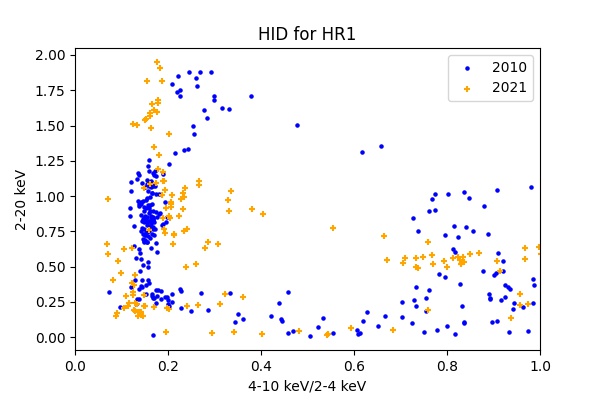}
    \includegraphics[height=0.2\textheight, width=.45\textwidth]{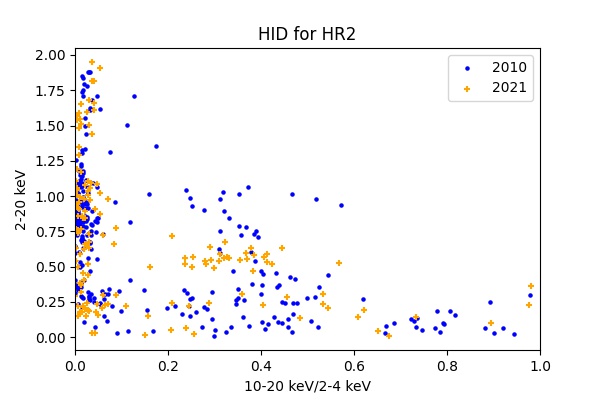}
    \caption{HIDs: a) HR1: energy bands 4-10 keV and 2-4 keV; b) HR2: energy bands 10-20 keV and 2-4 keV. Counts (flux) in energy range 2-20 keV on y axis of both plots.}
\label{fig:3}
\end{figure} 

\subsection{Hardness}
Hardness ratio (HR) is defined as the ratio between a high energy band and a low energy band. It is calculated by dividing the high energy band counts by the low energy band counts. The higher this ratio, the harder the spectrum. HID is obtained by summing the counts of multiple bands and plotting it against the hardness ratio. In most cases, the HID follows a 'q' shape, starting from the hard state at the right branch, transitioning to the soft state through the horizontal line on top, this horizontal part consisting of HIMS and SIMS, the left part being the soft state, returning back to the hard state through the bottom horizontal line.\\

Here we have used two different HRs, HR1(4-10 keV/2-4 keV) and HR2(10-20 keV/2-4 keV). For intensity, we have summed over the 2-20 keV range. HR2 does not provide good enough statistics, as we saw we had significantly less contribution in the energy range 10-20 keV, mostly around zero (thus resulting in an HR mostly equal to zero) (Fig. \ref{fig:3}). Therefore we concentrate on HR1 for our comparison.\\

In HR1, (Fig. \ref{fig:3}), many points are seen to be concentrated around lower values of HR, and the flux varies in the left branch. Three horizontal branches can be seen, the top one consisting of points from the 2010 observation, the middle one from the 2021 observation, and the bottom one consisting of both but dominated by the 2010 observation. 2010 outburst lasted longer, thus resulting in more data points on the HID. Considering the distribution percentage of the points, the 2021 outburst seems to be softer than the 2010 outburst.\\

From Fig. \ref{fig:2} we can see that the 2010 outburst ended with a harder flux than the end of the 2021 outburst. We can see a clear transition from LHS to HSS via HIMS and SIMS and then back to LHS, thus completing the 'q' shape. 


\subsection{Power Density Spectra}
I analyzed two orbits of data observed by LAXPC/AstroSat on 13\textsuperscript{th} February 2021, Orbit numbers 29094 and 29095, using HEASoft and LAXPCsoft. The Power Density Spectra (PDS) of both these orbits lacked the presence of Quasi Periodic Oscillations (QPOs). QPOs are an important characteristic of transient blackholes and are studied using the inspection of power density spectra (PDS). Low frequency QPOs are common in black hole transients in the HSS, HIMS, and SIMS and end of LHS \cite{belloni}. 

\section{Discussion and Results}\label{sec:filetypes}
From the lightcurve, we saw how the 2010 outburst lasted longer than the 2021 outburst, but the maximum flux went higher in the latter outburst. From the hardness ratio and HID, we saw that the 2010 outburst was harder. While the 2021 outburst started from a hard state, the 2010 outburst had an intermediate hardness ratio at the very begging, moved to the LHS, and then moved into HSS. Both the outbursts start off from the right end of the HID (refer to HR1), and move into the HIMS, here 2010 had higher flux but similar hardness. Then 2021 had a softer SIMS than 2010, also a brighter one (at the top left corner), both moved into HSS, moving back to LHS at the end, 2010 had a brighter HIMS than 2021 again. The two orbits of 2021 outburst data I analyzed did not possess any QPOs. In accordance with the LHS state, it appeared to be from the lightcurve and HR1. HR2 did not show much variation. Both the outbursts had most points concentrated around zero hardness ratio. Due to the lack of significant contribution in the 10-20 keV range, we chose to ignore inferring differences from HR2.

\begin{figure}
    \centering
    \includegraphics[height=0.2\textheight, width=.4\textwidth]{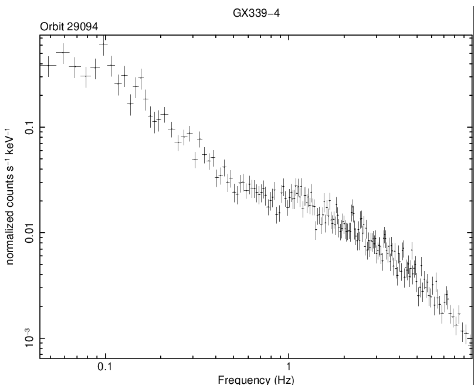}
    \includegraphics[height=0.2\textheight, width=.4\textwidth]{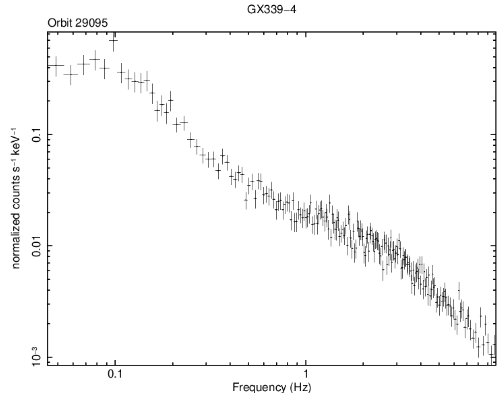}
    \caption{PDS: a) Orbit 29094; b) Orbit 29095. Frequency (Hz) on the x axis and Power on the y axis. Both of the plots lack presence of QPOs.}
\label{fig:4}
\end{figure}

\bibliographystyle{plain}
\bibliography{main}

\end{document}